# Einstein's 1916 derivation of the Field Equations

Galina Weinstein

24/10/2013

Abstract: In his first November 4, 1915 paper Einstein wrote the Lagrangian form of his field equations. In the fourth November 25, 1915 paper, Einstein added a trace term of the energy-momentum tensor on the right-hand side of the generally covariant field equations. The main purpose of the present work is to show that in November 4, 1915, Einstein had already explored much of the main ingredients that were required for the formulation of the final form of the field equations of November 25, 1915. The present work suggests that the idea of adding the second-term on the right-hand side of the field equation might have originated in reconsideration of the November 4, 1915 field equations. In this regard, the final form of Einstein's field equations with the trace term may be linked with his work of November 4, 1915. The interesting history of the derivation of the final form of the field equations is inspired by the exchange of letters between Einstein and Paul Ehrenfest in winter 1916 and by Einstein's 1916 derivation of the November 25, 1915 field equations.

In 1915, Einstein wrote the vacuum (matter-free) field equations in the form:[1]

$$(1)\ \frac{\partial \Gamma^{\sigma}_{\mu\nu}}{\partial x_{\alpha}} + \Gamma^{\sigma}_{\mu\beta}\Gamma^{\beta}_{\nu\alpha} = 0,$$

for all systems of coordinates for which $\sqrt{-g} = 1$.

It is sufficient to note that the left-hand side represents the gravitational field, with $g_{\mu\nu}$ the metric tensor field.

Einstein wrote the field equations in Lagrangian form. The action,[2]

$$(2)\ \delta\left\{\int L d\tau\right\} = 0,$$

and the Lagrangian,

$$(3)\ L = g^{\mu\nu}\Gamma^{\alpha}_{\mu\beta}\Gamma^{\beta}_{\nu\alpha}.$$

$$\sqrt{-g} = 1$$

Using the components of the gravitational field:

$$(4)\, \Gamma^{\tau}_{\mu\nu} = -\frac{1}{2} g^{\tau\alpha} \left( \frac{\partial g_{\mu\alpha}}{\partial x_{\nu}} + \frac{\partial g_{\nu\alpha}}{\partial x_{\mu}} - \frac{\partial g_{\mu\nu}}{\partial x_{\alpha}} \right)$$

Einstein wrote the variation:

$$(5)\, \delta L = -\Gamma^{\alpha}_{\mu\beta} \Gamma^{\beta}_{\nu\alpha} \delta g^{\mu\nu} - \Gamma^{\alpha}_{\mu\beta} \delta g^{\mu\beta}_{\alpha},$$

which gives:[3]

$$(6)\, \frac{\partial L}{\partial g^{\mu\nu}} = -\Gamma^{\alpha}_{\mu\beta} \Gamma^{\beta}_{\nu\alpha}, \qquad \frac{\partial L}{\partial g^{\mu\nu}_{\sigma}} = \Gamma^{\sigma}_{\mu\nu}.$$

We now come back to (2), and we have,

$$(7)\, \frac{\partial}{\partial x_{\alpha}} \left( \frac{\partial L}{\partial g^{\mu\beta}_{\alpha}} \right) - \frac{\partial L}{\partial g^{\mu\nu}} = 0.$$

Inserting (6) into (7) gives the field equations (1). Hence, the vacuum field equations can be written in the form (7).

On November 4, 1915 Einstein presented the first version of his 1915 general theory of relativity, "On the General Theory of Relativity". He wrote the field equations with the sources (the energy-momentum tensor, "matter") on the right-hand side of (7):[4]

$$(8)\, \frac{\partial}{\partial x_{\alpha}} \left( \frac{\partial L}{\partial g^{\mu\beta}_{\alpha}} \right) - \frac{\partial L}{\partial g^{\mu\nu}} = -\kappa T_{\mu\nu}.$$

In the first November 4, 1915 paper Einstein adopted $\sqrt{-g} = 1$ as a postulate. Hence, (8) were not the generally covariant Einstein field equations. A week later, on November 11, 1915, in an "Addendum" to the first paper, Einstein dropped this postulate and adopted it as a coordinate condition $\sqrt{-g} = 1$.

Inserting (6) into (8), the November 4, 1915 field equations are

$$(9)\, \frac{\partial \Gamma^{\sigma}_{\mu\nu}}{\partial x_{\alpha}} + \Gamma^{\sigma}_{\mu\beta} \Gamma^{\beta}_{\nu\alpha} = -\kappa T_{\mu\nu}.$$

Einstein multiplied (8) by $g^{\mu\nu}_{\sigma}$ with summation over the indices $\mu$ and $\nu$ and obtained:[5]

$$(10) \quad \frac{\partial}{\partial x_\alpha}\left(g_\sigma^{\mu\nu}\frac{\partial L}{\partial g_\alpha^{\mu\nu}}\right) - \frac{\partial L}{\partial x_\sigma} = -\kappa T_{\mu\nu}g_\sigma^{\mu\nu},$$

He defined the energy tensor of the gravitational field:

$$(11) \quad -2\kappa t_\sigma^\alpha = g_\sigma^{\mu\nu}\frac{\partial L}{\partial g_\alpha^{\mu\nu}} - \delta_\sigma^\alpha L.$$

Equation (11) expresses the law of conservation of momentum and energy for the gravitational field. The quantities $t_\sigma^\alpha$ are the energy components of the gravitational field.

With the second of equations (6), the components of the gravitational field, equation (11) may be written as follows,[6]

$$(12) \quad \kappa t_\sigma^\alpha = \frac{1}{2}\delta_\sigma^\alpha g^{\mu\nu}\Gamma_{\mu\beta}^\alpha \Gamma_{\nu\alpha}^\beta - g^{\mu\nu}\Gamma_{\mu\beta}^\alpha \Gamma_{\nu\sigma}^\beta.\text{[7]}$$

Einstein multiplied (9) by $g^{\nu\sigma}$ and summed over $\nu$:[8]

$$(13) \quad g^{\nu\sigma}\frac{\partial \Gamma_{\mu\nu}^\alpha}{\partial x_\alpha} - g^{\nu\sigma}\Gamma_{\nu\beta}^\alpha \Gamma_{\nu\alpha}^{\beta\lambda} = -\kappa g^{\nu\sigma}T_{\mu\nu} = -\kappa T_\mu^\tau.$$

The second term on the left-hand side above is the second term on the right-hand side of (12), and so Einstein combined between the two and obtained,

$$(14) \quad \frac{\partial}{\partial x_\alpha}\left(g^{\sigma\beta}\Gamma_{\mu\beta}^\alpha\right) - \frac{1}{2}\delta_\sigma^\alpha g^{\mu\nu}\Gamma_{\mu\beta}^\lambda \Gamma_{\nu\lambda}^\beta = -\kappa\left(T_\mu^\tau + t_\mu^\lambda\right).$$

In a letter to Paul Ehrenfest and later in his 1916 paper, "The Foundation of the General Theory of Relativity", Einstein defined:[9]

$$(15) \quad g^{\mu\nu}\Gamma_{\mu\beta}^\lambda \Gamma_{\nu\lambda}^\beta = \kappa t.$$

Inserting (15) in the November 4, 1915 equation (14), we have,

$$(16) \quad \frac{\partial}{\partial x_\alpha}\left(g^{\sigma\beta}\Gamma_{\mu\beta}^\alpha\right) - \frac{1}{2}\delta_\mu^\alpha \kappa t = -\kappa\left(T_\mu^\sigma + t_\mu^\sigma\right).$$

Einstein rewrote equation (16) in a form valid for matter-free gravitational fields:[10]

$$(17) \quad \frac{\partial}{\partial x_\alpha}\left(g^{\sigma\beta}\Gamma^\alpha_{\mu\beta}\right) = -\kappa\left(t^\sigma_\mu - \frac{1}{2}\delta^\sigma_\mu t\right), \quad \sqrt{-g} = 1.$$

Here the November 4, 1915 postulate is adopted as a coordinate condition $\sqrt{-g} = 1$. Einstein was then able to rewrite equation (17) as follows:[11]

$$(18) \quad \frac{\partial}{\partial x_\alpha}\left(g^{\sigma\beta}\Gamma^\alpha_{\mu\beta}\right) = -\kappa\left[\left(t^\sigma_\mu + T^\sigma_\mu\right) - \frac{1}{2}\delta^\alpha_\mu(t + T)\right], \quad \sqrt{-g} = 1.$$

Einstein gave an interpretation of this equation to Ehrenfest,[12]

"This equation is interesting, because it shows that the source of the gravitation lines is determined solely by the sum $T^\nu_\sigma + t^\nu_\sigma$, as indeed it must be expected".

Compare between equations (16) and (18):

$$(16) \quad \frac{\partial}{\partial x_\alpha}\left(g^{\sigma\beta}\Gamma^\alpha_{\mu\beta}\right) = -\kappa\left[\left(t^\sigma_\mu + T^\sigma_\mu\right) - \frac{1}{2}\delta^\alpha_\mu t\right].$$

We now see, from the above derivation that, sometime before November 25, 1915, it appears as if Einstein added to equation (16) the term: $-\frac{1}{2}\delta^\alpha_\mu T$, and this enabled him to write equation (18). The added term just mentioned, though more than just and additional term, is $T = T^\mu_\mu$ "Laue's scalar".[13] In fact, Einstein wrote Ehrenfest about the "'inevitability' requirement for the additional term $-1/2\ g_{im}T$".[14]

Taking into account (17) and (1) and multiplying (18) by $g_{\mu\nu}$ leads to the field equations of November 25, 1915 valid for a coordinate system $\sqrt{-g} = 1$:[15]

$$(19) \quad \frac{\partial \Gamma^\alpha_{\mu\nu}}{\partial x_\alpha} + \Gamma^\alpha_{\mu\beta}\Gamma^\beta_{\nu\alpha} = -\kappa\left(T_{\mu\nu} - \frac{1}{2}g_{\mu\nu}T\right), \quad \sqrt{-g} = 1.$$

On November 28, 1915, Einstein wrote Arnold Sommerfeld about the theory which can be simplified by:[16] "choosing the frame of reference so that $\sqrt{-g} = 1$. Then the equations take on the form,

$$-\sum_l \frac{\partial \begin{Bmatrix} im \\ l \end{Bmatrix}}{\partial x_l} + \sum_{\alpha\beta} \begin{Bmatrix} i\alpha \\ \beta \end{Bmatrix} \begin{Bmatrix} m\beta \\ \alpha \end{Bmatrix} = -\kappa \left( T_{im} - \frac{1}{2} g_{im} T \right)$$

I had considered these equations with [Marcel] Grossman already 3 years ago, up to the second term on the right-hand side".

"These equations […] up to the second term on the right-hand side" are equations (9) of the November 4, 1915 paper.

---

[1] Einstein, Albert, "Die Grundlage der allgemeinen Relativitätstheorie", *Annalen der Physik* 49, 1916, pp. 769-822; p. 803; Einstein, Albert (1915b), "Die Feldgleichungen der Gravitation", *Königlich Preußische Akademie der Wissenschaften* (Berlin). *Sitzungsberichte*, 1915, pp. 844-847; p. 832.

[2] Einstein, 1916, p. 804; Einstein, Albert (1915a), "Zur allgemeinen Relativitätstheorie", *Königlich Preußische, Akademie der Wissenschaften* (Berlin). *Sitzungsberichte*, 1915, pp. 778-786; p. 784. See my paper: Weinstein, Galina, "FROM THE BERLIN "ENTWURF" FIELD EQUATIONS TO THE EINSTEIN TENSOR II: November 1915 until March 1916", ArXiv: 1201.5353v1 [physics.hist-ph], 25 January, 2012, pp. 39-52. Einstein first defined: $g_\sigma^{\mu\nu} = \frac{\partial g^{\mu\nu}}{\partial x_\sigma}$.

All symbols used here play the same part as they do in Einstein's original 1916 and 1915 works, and thus this is a historical derivation.

[3] Einstein, 1916, p. 805; Einstein, 1915a, p. 784.

[4] Einstein, 1915a, p. 784.

[5] Einstein, 1916, p. 805; Einstein, 1915a, p. 784. Using the definition $\frac{\partial g_\sigma^{\mu\nu}}{\partial x_\alpha} = \frac{\partial g_\alpha^{\mu\nu}}{\partial x_\sigma}$ Einstein wrote:

$$g_\sigma^{\mu\nu} \frac{\partial}{\partial x_\alpha} \left( \frac{\partial L}{\partial g_\alpha^{\mu\nu}} \right) = \frac{\partial}{\partial x_\alpha} \left( g_\sigma^{\mu\nu} \frac{\partial L}{\partial g_\alpha^{\mu\nu}} \right) - \frac{\partial L}{\partial g_\alpha^{\mu\nu}} \frac{\partial g_\alpha^{\mu\nu}}{\partial x_\sigma}$$

[6] Einstein, 1916, p. 806; Einstein, 1915a, p. 785.

[7] Einstein used: $\frac{\partial g_{\mu\nu}}{\partial x_\sigma} = -g^{\mu\tau} \begin{Bmatrix} \tau\sigma \\ \nu \end{Bmatrix} - g^{\nu\tau} \begin{Bmatrix} \tau\sigma \\ \mu \end{Bmatrix}$ when writing equation (12).

[8] Einstein, 1915a, p. 785.

[9] Einstein to Ehrenfest, January 24 or later, 1916, *The Collected Papers of Albert Einstein, Vol. 8 (CPAE, Vol. 8): The Berlin Years: Correspondence, 1914–1918*, Schulmann, Robert, Kox, A.J., Janssen, Michel, Illy, Jószef (eds.), Princeton: Princeton University Press, 2002, Doc. 185. In the letter to Ehrenfest Einstein wrote (15): $g^{\mu\nu} \Gamma_{\mu\beta}^\lambda \Gamma_{\nu\lambda}^\beta = -\kappa \left( t_\mu^\sigma - \frac{1}{2} \delta_\mu^\sigma t \right)$.

[10] Einstein, 1916, p. 806.

[11] Einstein, 1916, p. 807.

[12] Einstein to Ehrenfest, January 24 or later, 1916, *CPAE*, Vol. 8, Doc. 185.

[13] Einstein, 1916, p. 808.

[14] Einstein to Ehrenfest, January 17, 1916, *CPAE*, Vol. 8, Doc. 182.

[15] Einstein, 1916, p. 808.

[16] Einstein to Sommerfeld, November 28, 1915, *CPAE*, Vol. 8, Doc. 153 (before inventing the summation convention in 1916).